\documentclass{article}
\usepackage{a4,amssymb,amsmath,amsxtra,epsfig}

\newcommand{\diff}{{\rm{d}}}
\newcommand{\N}{{\mathcal{N}}}
\newcommand{\CF}{{\mathcal{C}}}

\begin{document}

\title{PCAC and the Deficit of Forward Muons \\
   in $\pi^+$   Production by Neutrinos}

\author{D.~Rein  and L.~M.~Sehgal}
\date{\it III. Physikalisches Institut and Institute of Theoretical Physics (E), RWTH Aachen \\ D-52056 Aachen, Germany}

\maketitle 

\begin{abstract}

The K2K experiment, using a fine-grained detector in a neutrino beam of energy 
$<E> \,\, \sim 1.3 \, \mathrm{GeV}$
has observed two-track events that can be interpreted as a coherent reaction 
$\nu_\mu + \N \to \mu^- + \N + \pi^+ \,\,(\N = \rm{C}^{12})$
or an incoherent process $\nu_\mu + (p,n) \to \mu^- + \pi^+ + (p,n)$, 
the final nucleon being unobserved. The data show a significant
deficit of forward-going muons in the interval 
$Q^2 \lesssim 0.1 \rm{GeV}^2$,
where a sizeable coherent signal is expected. 
We attempt an explanantion of this effect, using a PCAC formula 
that includes the effect of the non-vanishing muon mass.
A suppression of about 25 \% is caused by a destructive interference of the axial 
vector and pseudoscalar (pion-exchange) 
amplitudes. The incoherent background is also reduced by 10 - 15 \%. As a consequence the discrepancy between theory and observation is significantly reduced. 

\end{abstract}

The K2K experiment has studied interactions of a low energy neutrino beam 
($<E> \,\, \sim 1.3 \, \rm{GeV}$)
in a fine-grained detector, designed as the ``near detector'' of a 
long-base-line neutrino oscillation experiment 
\cite{Hasegawa:2005td,Nakaya:2005gz,Yokoyama:2004}. 
Evidence has been found for two-track events, which can be interpreted
as either 
$\nu_\mu + \N \to \mu^{-} + \N + \pi^+ $
(coherent $\pi^+$ production on a nuclear target) or incoherent
$\pi^+$ production 
$\nu_\mu + (p,n) \to \mu^{-} + \pi^+ + (p,n)$, 
where the final nucleon is unobserved. The data have been compared with 
simulations based on a model for coherent $\pi^0$ 
production \cite{Rein:1982pf}, 
and a model for incoherent single pion production via nucleon
resonances
\cite{Rein:1980wg}.
It is stated that in comparison with the simulations, 
the two-track data show ``a significant deficit of forward-going muons''
in the kinematic interval
$Q^2 \lesssim 0.1 \rm{GeV}^2$,
in which a sizeable coherent contribution is expected.
In this Letter, we examine a possible explanation of this effect. 

As is well-known, neutrino scattering in the forward-scattering configuration
is described by Adlers PCAC theorem \cite{Adler:1964yx}. 
For any inelastic charged current reaction 
$\nu_\mu + N \to \mu^{-} + F$, 
where $F$ denotes an inelastic channel, the cross section, 
neglecting the muon mass, is
\begin{equation}
\label{CS-PCAC1}
    \Big( \frac{\diff \sigma}{\diff x\,\diff y} \Big)_{\rm{PCAC}} 
    = \frac{G^2 M E}{\pi^2} f_{\pi}^2 (1-y) 
    \sigma (\pi^+ + N \to F) \bigg|_{E_\pi = Ey}
\end{equation}
where $x = Q^2 / 2 M E y$ and $y = \nu / E$, $\nu$ 
being the energy transfer and $E$ the neutrino energy. 
The pion decay constant has the value 
$f_\pi \approx 0.93 m_\pi$
and $M$ denotes the nucleon mass. 
The extrapolation of the PCAC formula to non-forward angles is given
by a slowly varying form-factor 
$[m_A^2 / (m_A^2 + Q^2)]^2$, 
with $m_A \approx 1 \rm{GeV}$.

There is an important modification of Eq. (\ref{CS-PCAC1}) when the
mass of the muon is taken into account. 
This modification can be found in a recent comment by
Adler \cite{Adler:2005ad}, and can be expressed as a simple
multiplicative correction factor 
\begin{equation}
\label{def-CF}
\CF = \big( 1 - \frac{1}{2} \frac{Q^2_{\rm{min}}}{Q^2 + m_\pi^2}
\big)^2
+ \frac{1}{4} y \frac{Q^2_{\rm{min}} (Q^2 - Q^2_{\rm{min}})}{(Q^2 +
  m_\pi^2)^2} 
\end{equation}
where
\begin{equation}
Q^2_{\rm{min}} = m_l^2\,\frac{y}{1-y} . 
\end{equation}
The range of the variable $Q^2$ is
\begin{equation}
Q^2_{\rm{min}} \le Q^2 \le 2\,M\,E\,y_{\rm{max}}, 
\end{equation}
where $y$ lies between $y_{\rm{min}} = m_\pi / E$ 
and $y_{\rm{max}} = 1 - m_l / E$. 
Thus the corrected PCAC formula, valid for small angle scattering, is 
\begin{eqnarray}
\label{CS-PCAC2}
    \Big( \frac{\diff \sigma}{\diff x\,\diff y} 
    \Big)_{{\rm{PCAC}},\,m_l \ne 0} 
    &=& \frac{G^2 M E}{\pi^2} f_{\pi}^2 (1-y) 
    \sigma (\pi^+ + \N \to F) \bigg|_{E_\pi = Ey} \nonumber\\
	&\cdot& \CF\,
    \theta(Q^2 - Q^2_{\rm{min}})\,
    \theta(y - y_{\rm{min}})\,
    \theta(y_{\rm{max}} - y)
\end{eqnarray}

The physical interpretation of the correction factor $\CF$ is as follows:
When the muon mass is not neglected, the reaction 
$\nu_\mu + \N \to \mu^{-} + F$
receives a contribution from the
exchange of a charged pion between the lepton vertex 
$\nu \to \mu^-$
and the hadron vertex
$\N \to F$. 
The coupling at the lepton vertex is 
$f_\pi\,m_l\,\bar{u}_\mu\,\gamma_5\,u_\nu$, and the amplitude
contains the characteristic pion propagator 
$(Q^2 + m_\pi^2)^{-1}$. 
This so-called pseudoscalar amplitude interferes with the 
remaining amplitude, which is free of the pion singularity, 
and which for $m_l = 0$ becomes proportional to 
$\big< F \big| \partial_\alpha A_\alpha \big| N \big>$
in the forward scattering configuration.
(For brevity we call this pole-free contribution the ``axial'' amplitude.) 
These two amplitudes interfere destructively. The destructive nature
of the interference is visible in the first term of the correction
factor $\CF$, which reads 
$\big( 1 - \frac{1}{2}\,\frac{Q^2_{\rm{min}}}{Q^2 + m_\pi^2} \big)^2$.
The two terms within the parentheses represent the axial and 
pseudoscalar amplitudes. The minus sign represents destructive interference.
This, in our opinion, is a partial explanation of the low-$Q^2$ 
deficit observed in the charged current $\pi^+$ production at low energies.

To see the impact of the correction factor $\CF$ we apply the modified
PCAC formula (\ref{CS-PCAC2}) to the coherent process 
$\nu_\mu + \N \to \mu^- + \N + \pi^+$, 
using the same model for nucleon coherence used in describing coherent 
$\pi^0$ production \cite{Rein:1982pf}. 
This cross section has the form 
\begin{equation}
\label{CS-2}
    \Big( \frac{\diff \sigma^{\pi^+}}{\diff x\,\diff y\,\diff|t|} 
    \Big)
    = 2\, 
    \Big( \frac{\diff \sigma^{\pi^0}}{\diff x\,\diff y\,\diff|t|} 
    \Big)
    \cdot \CF\,
    \theta(Q^2 - Q^2_{\rm{min}})\,
    \theta(y - y_{\rm{min}})\,
    \theta(y_{\rm{max}} - y),
\end{equation}
where 
$\diff \sigma^{\pi^0} / \diff x\,\diff y\,\diff|t|$
is given explicitly in Eq.(10) of Ref.\cite{Rein:1982pf}. 

Replacing $x$ by $Q^2/(2M E y)$ and integrating over the 
variables $t$ and $y$ we
obtain the $Q^2$-distribution 
$\diff \sigma / \diff Q^2$.
For the purpose of understanding the essential origin of the
$Q^2$ suppression, it is enough to treat the
pion-nucleon cross-section and the effects of
nucleon absorption as constants (a complete calculations confirms our conclusion).  The integration over
the variable $t$ then yields the approximate result
\begin{equation}
    \Big( \frac{\diff \sigma}{\diff y\,\diff Q^2} 
    \Big)
    \sim \frac{1 - y}{y}\,\exp [-b |t|_{\rm{min}}]\,\CF,
\end{equation}
where $|t|_{\rm{min}}$ is given by
\begin{equation}
    |t|_{\rm{min}} = 2 \,E^2\,y^2\,
\bigg\{ 
1 + \frac{M x}{E y} -
\frac{m_\pi^2}{2\,E^2\,y^2} - \sqrt{1 + \frac{2 M x}{E y}}\,
\sqrt{1 - \frac{m_\pi^2}{E^2\,y^2}} 
\bigg\}.
\end{equation}
Integration over $y$ produces the distribution $\diff \sigma / \diff Q^2$
shown in Fig.~1, which clearly exhibits a low-$Q^2$ suppression in the
region $Q^2 < 0.1 {\rm GeV}^2$ amounting to about $< \CF_{coh}> \approx 0.75$ at $E_{\nu} = 1.3$ GeV. The empirical $Q^2$-distribution includes an incoherent background from $\nu_\mu + (p,n) \to \mu^- + \pi^+ + (p,n)$. Using the resonance model of Ref. [5], and introducing the suppression factor $\CF$ in the scalar part of the cross section, we estimate an effective resonant suppression factor $< \CF_{res}> \approx 0.85 - 0.90$. As a consequence, the discrepancy between theory and K2K observation is reduced to about $2 \sigma$ \cite{Sehgal}. 

Our explanation of the low-$Q^2$ deficit implies that the 
effect occurs only for charged current scattering, where the muon mass plays a role. 
The fact that the muon mass is of the same order as the pion mass
appearing in the pion-propagator is important.
The effect diminishes with increasing neutrino
energy. 
The neutral current channels are unaffected.
In this connection it is significant that the recent K2K measurement of
$\nu_\mu + \N \to \nu_{\mu} + \N + \pi^0$ \cite{Nakayama:2004dp}
finds that the angular distribution and the momentum spectrum of the 
$\pi^0$ are in good agreement with the Monte Carlo expectations 
\cite{Rein:1982pf,Rein:1980wg}.

It is to be hoped that the new high resolution detectors such as SciBoone and Minerva 
will be able to test these ideas concerning the forward-muon deficit in an incisive way.
We draw attention to some recent papers \cite{SinghKartavsevAlvarezRuso} that may have a bearing on the subject of this paper.

Acknowledgement: We are indebted to Dr. Yuichiro Kiyo, 
Volker Schulz and Georg Kreyerhoff for their kind assistance in the preparation of the manuscript.
One of us (L.M.S.) wishes to thank J. Morfin and the organizers of NuInt07 for the invitation to present these ideas at their Workshop in Fermilab, May 30 - June 3, 2007.

\newpage

\begin{figure}
\begin{center}
\epsfxsize=10cm
\epsffile{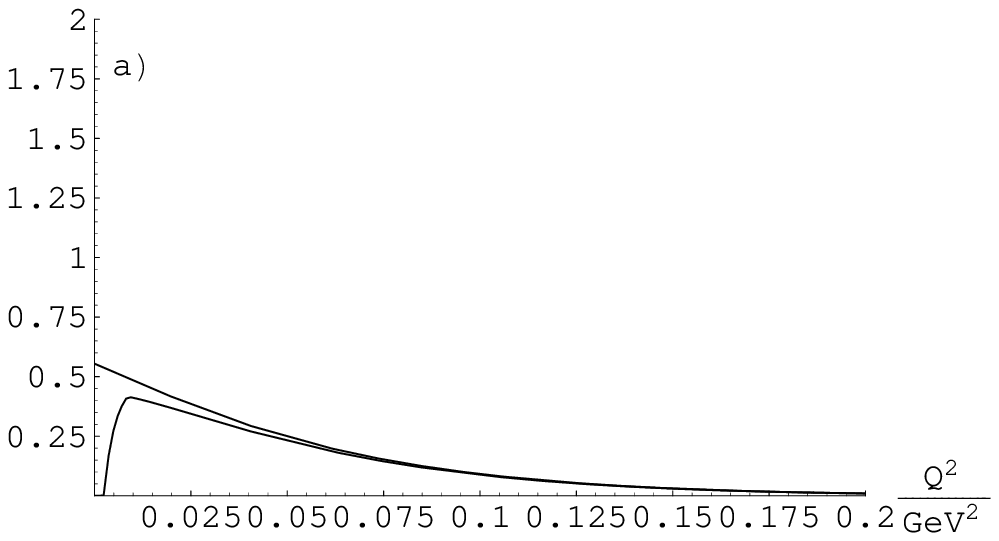}
\epsfxsize=10cm
\epsffile{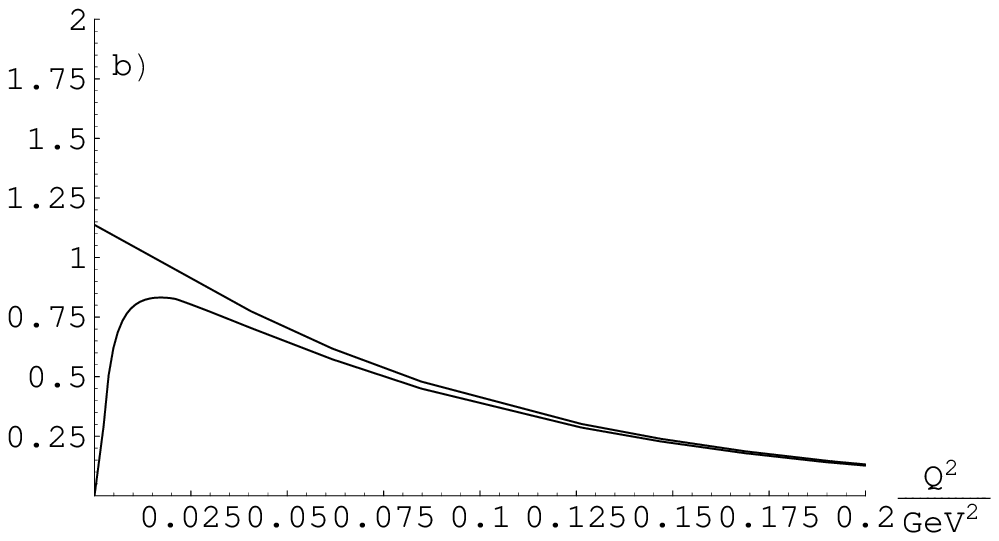}
\epsfxsize=10cm
\epsffile{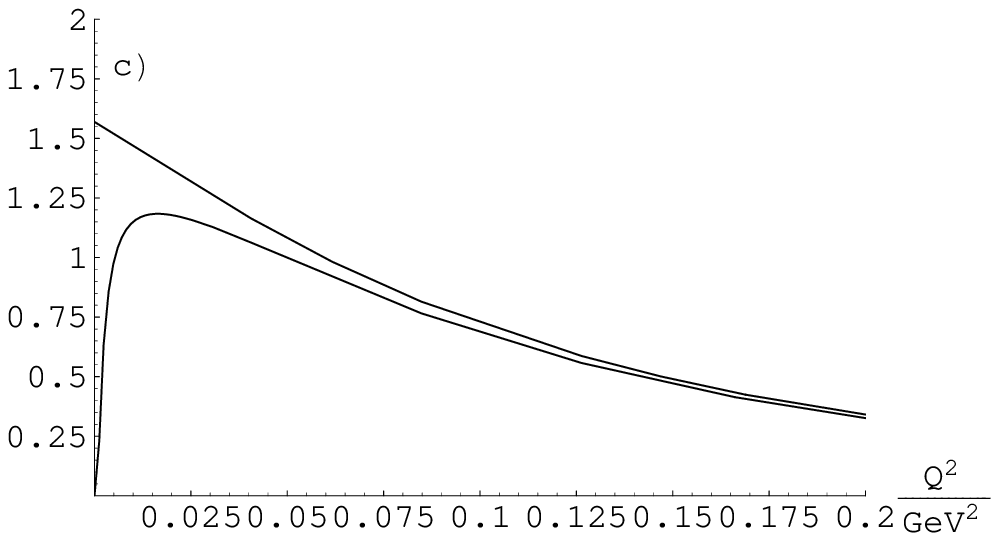}
\caption{Effect of muon-mass correction on $Q^2$-dependence
of coherent $\pi^+$ production. Curves are derived from
the simplified expression given in Eq.~(8). The target mucleus is
Carbon, and the neutrino energy is (a) 0.8 GeV, (b) 1.3 GeV and
(c) 2.0 GeV. In each figure, the upper (lower) curve corresponds
to the case $m_l =0$ ($m_l\neq 0$) \label{figure1}}
\end{center}
\end{figure}

\end{document}